# Antiferromagnetic Piezospintronics

Zhiqi Liu*, Zexin Feng, Han Yan, Xiaoning Wang, Xiaorong Zhou, Peixin Qin, Huixin Guo, Ronghai Yu, Chengbao Jiang

School of Materials Science and Engineering, Beihang University, Beijing 100191, China

*email: zhiqi@buaa.edu.cn

**Antiferromagnets naturally exhibit three obvious advantages over ferromagnets for memory device applications: insensitivity to external magnetic fields, much faster spin dynamics (~THz) and higher packing density due to the absence of any stray field. Recently, antiferromagnetic spintronics emerges as a cutting-edge field in the magnetic community. The key mission of this rapidly rising field is to steer the spins or spin axes of antiferromagnets via external stimuli and then realize advanced devices based on their physical property changes. Herein, the state of the art of antiferromagnetic spintronics is presented. Subsequently, the history of ferromagnetic/ferroelectric multiferroic composites is briefly revisited. Finally, we introduce an ultralow-power, long-range, and magnetic-field-insensitive approach for harnessing antiferromagnetic spins based on our recent experimental progress, *i.e.*, piezoelectric strain control. Relevant theoretical and experimental studies have formed an attractive new branch in antiferromagnetic spintronics, which we coin as antiferromagnetic piezospintronics.**

1. **Some important steps in antiferromagnetic spintronics**

In antiferromagnets, the magnetic moments of two or more atomic sublattices are ordered in such a way that the net magnet moment is zero. As a result, antiferromagnetic materials have been mainly used so far as auxiliary materials in spin valve structures for magnetic recording, serving as exchange-bias pinning layers for adjacent ferromagnetic films to modify their magnetic switching. However, this situation may change in the future.

Compared with ferromagnets for which the stability of spins is characterized the anisotropy field $H_A$, the spin-flop field $H_{SF}$, which represents the robustness of spins of an antiferromagnet under external magnetic fields and involves both the antiferromagnetic exchange field $H_E$ and the anisotropic field $H_A$ via $H_{SF} \approx \sqrt{2H_E H_A}$, is typically much larger than the $H_A$ of ferromagnets and can reach several tens T or even more than 100 T. Similarly, the zero-field resonance frequency $\omega_{AFM}$ of an antiferromagnet involves the antiferromagnetic exchange field $H_E$ as well due to spin canting via $\omega_{AFM} \approx r\sqrt{2H_E H_A} \approx rH_{SF}$, where $r$ is the gyromagnetic ratio of an electron. It can be three orders of magnitude higher than that of ferromagnets $\omega_{FM} \approx rH_A$ (typically GHz) and reaches THz. For example, the study on the laser-induced spin reorientation in antiferromagnetic $TmFeO_3$ in 2004 shows that the antiferromagnetic spins can be manipulated on a timescale of a few picoseconds.[1]

In 2006, Núñes *et al.* proposed a pioneering theory that spin transfer torques can induce the order parameter orientation switching in antiferromagnetic metals, which is well similar to the ferromagnetic case.[2] However, they pointed out that compared with the ferromagnetic case, the critical current for antiferromagnetic order parameter switching can be smaller because of the absence of shape anisotropy and also because spin torques can act through the entire volume of an antiferromagnet. On the other hand, as the magnetic order in an antiferromagnet is staggered, only correspondingly staggered torques can drive coherent order parameter switching. Soon in 2007, different experimental groups demonstrated that the exchange bias of a ferromagnet/antiferromagnet bilayer system can be altered by a current and thus provided indirect evidences for current-induced torques in antiferromagnetic metals.[3-5] Subsequently, Gomonay and Loktev proposed the phenomenological model that describes the spin transfer torques in antiferromagnets.[6,7] These early studies were summarized by MacDonald and Tsoi in the review paper that emphasizes the concept of antiferromagnetic metal spintronics[8] and also by Gomonay and Loktev in the review paper that emphasizes spintronics of antiferromagnetic systems from a theoretical point of view.[9]

In 2011, Park *et al.* creatively reversed the stacking order of the antiferromagnetic layer IrMn and the ferromagnetic layer NiFe in a spin-valve-like tunnel junction structure, where the antiferromagnetic IrMn served as the key functional layer for generating tunnel anisotropic magnetoresistance, while the ferromagnetic NiFe layer was utilized to rotate the antiferromagnetic spin axis of IrMn via the interfacial exchange spring effect.[10] Surprisingly, a more than 100% tunneling anisotropic magnetoresistance was achieved at 4 K. This device proves that antiferromagnetic materials could work as pivotal components in spintronic devices instead of simply serving as pinning layers for ferromagnetic materials and is thus an important step towards antiferromagnetic spintronics. Nevertheless, as the switching of antiferromagnetic spins is realized by the magnetization switching of the ferromagnetic layer, both the switching speed and magnetic-field response of such a device are dominated by the ferromagnetic layer. Thus, this device does not possess any advantage of antiferromagnets over ferromagnets. In addition, the tunneling anisotropic magnetoresistance decayed rapidly with increasing temperature and consequently could not work at high temperatures above 100 K.

In 2014, Marti *et al.* demonstrated the first room-temperature antiferromagnetic memory in FeRh, an intermetallic alloy exhibiting a first-order magnetic phase transition at ~365 K from a low-temperature antiferromagnetic phase to a high-temperature ferromagnetic phase.[11] They applied a magnetic field of 9 T during cooling from the ferromagnetic phase at 400 K to the antiferromagnetic phase at 300

K, which is to stabilize a metastable spin axis perpendicular to the applied magnetic field at room-temperature. By changing the cooling field direction by 90° at 400 K, the antiferromagnetic spin axis is accordingly rotated by 90° at room temperature as well. As a result, high- and low-resistance states were achieved by virtue of the anisotropic magnetoresistance in antiferromagnets. This work illustrates a clear picture that an antiferromagnetic alone can work as a memory, which largely promotes the development of the antiferromagnetic spintronics. Nevertheless, the information writing process in such an antiferromagnetic memory device needs both thermal heating and a large magnetic field to switch the antiferromagnetic order parameter and hence it is not really easy for practical memory device application yet.

Meanwhile, Železný *et al*. theoretically predicted the staggered Néel-order spin-orbit torque fields in 2014, whose sign alternates between the spin sublattices, can be induced by an electric current in antiferromagnetic $Mn_2Au$ and are able to trigger ultrafast spin-axis reorientation.[12] It is similar to the relativistic spin-orbit torques in ferromagnets[13] with broken bulk or structural inversion symmetry but different from the spin transfer torques which were proposed for antiferromagnets in early reports.[2,6,7] Later in 2016, the Néel spin-orbit torque switching of the antiferromagnetic spin axis was experimentally achieved by the same group led by Prof. Jungwirth in tetragonal antiferromagnetic semiconductor CuMnAs,[14] which marks the birth of an electrically-controlled antiferromagnetic memory. The current density for switching the antiferromagnetic spin axis in CuMnAs was $4 \times 10^6$ $A/cm^2$, and is much smaller than that in ferromagnetic metals. Furthermore, the microelectronic compatibility[15] and the THz electrical writing speed have been demonstrate for the CuMnAs-based antiferromagnetic memory.[16] In addition, the Néel spin-orbit torque driven switching has been also realized in an antiferromagnetic metal $Mn_2Au$ as theoretically predicted,[17] which is capable of reducing Joule heating due to the lower resistivity relative to semiconducting CuMnAs.

Additionally, some exotic effects have been also revealed for antiferromagnets in recent years. For example in 2014, Mendes *et al*. experimentally discovered the large inverse spin Hall effect in non-collinear antiferromagnetic metal $Mn_{80}Ir_{20}$[18] and Zhang *et al*. uncovered spin Hall effects in metallic collinear antiferromagnets including CuAu-I-type MnPt, MnIr, MnPd and MnFe.[19] In the same year, Chen *et al*. theoretically proposed spin pumping and spin-transfer torques in antiferromagnets.[20] Two years later in 2016, Wu *et al*. reported the experimental observation of the spin Seebeck effect in antiferromagnetic $MnF_2$.[21] These exciting progresses on spin current-related phenomena in antiferromagnets have been summarized in some nice review articles by Gomonay *et al*.,[22] Jungfleisch *et al*.[23] and Sklenar *et al*.[24] More recently, magnetism-related spin Hall and inverse spin Hall effects have been experimentally observed by Kimata *et al*. in non-collinear antiferromagnet – $Mn_3Sn$, which are coined as magnetic spin Hall effect and magnetic inverse spin Hall effect, respectively.[25]

Motivated by the above works and other relevant studies[26-32], antiferromagnetic spintronics has begun to take shape and there are a surge of theoretical and experimental studies in this field in recent years.[33,34] Moreover, *Nature Physics* published a collection of reviews and commentaries[35-40] focusing specifically on antiferromagnetic spintronics in March of 2018 and highlighted them with an editorial article entitled "Upping the anti",[41] which emphasizes that antiferromagnetic spintronics can create exotic forms of magnetism into practical applications.

2. **Piezoelectric control of ferromagnetism in multiferroic heterostructures**

On the other hand, multiferroic composites/heterostructures consisting of ferromagnetic and ferroelectric materials are multifunctional materials for smart materials applications. The initial idea came as early as in 1972 when van Suchtelen proposed the magnetoelectric effect in a composite material with one magnetostrictive phase and one piezoelectric phase and further demonstrated such an effect in a $BaTiO_3$-$CoFe_2O_4$ eutectic composite,[42] which, by the way, is consistent with theoretical calculations based on the Green's function method and the perturbation theory.[43] In this new class of composites, one can apply a magnetic field to induce a distortion of the magnetostrictive phase, which subsequently deforms the piezoelectric phase and then generates an electric field, or vice versa.

Due to the large application potential of this new class of composites as actuators, sensors, microelectronic mechanical systems, switches, and memory devices,[44,45] laminate composites[46,47] and thin-film heterostructures gradually appear to be centers of attention, which make this field extraordinarily active.[48-58] An important aspect about ferromagnetic/ferroelectric heterostructures for memory applications is the ultralow energy consumption. When one conveniently uses electric-field-generated strain from the ferroelectric material to modify the magnetic and/or electrical properties of the ferromagnetic materials, Joule heating is largely suppressed because of the highly insulating nature of ferroelectric materials, which can potentially lead to extremely small power consumption during information writing, aJ/bit.[58] Thus, piezoelectric strain control provides a superior approach for the contemporary aJ information technique. In addition, giant magnetoelectric coefficients $\alpha$ have been achieved in recent years, such as $Co_{40}Fe_{40}B_{20}$/PMN-PT ($\alpha \sim 10^{-6}$ s/m)[59] and FeRh/$BaTiO_3$ ($\alpha \sim 10^{-5}$ s/m).[60,61]

It is worth mentioning that the piezoelectric strain triggered from ferroelectric or piezoelectric materials by electric fields is dynamic, reversible and works in a long range manner. For example, Biegalski *et al*. found that the piezoelectric strain is 100% transferred from PMN-PT substrates to 200-nm-thick oxide films that are epitaxially grown on PMN-PT.[62] Moreover, Dekker *et al*. found that the reversible in-plane piezoelectric strain in a 600-nm-thick $La_{0.7}Sr_{0.3}MnO_3$/$SrTiO_3$ superlattice containing 100 oxide interfaces is vertically homogeneous.[63] The piezoelectric strain are in sharp contrast to the epitaxial strain owing to the lattice mismatch in epitaxial growth, which is static, irreversible and relaxes quickly from interfaces.[64-67]

3. **Antiferromagnetic Piezospintronics**

In this final section, we would like to link antiferromagnetic spintronics with piezoelectric strain control and hence introduce the formation of a new subfield – antiferromagnetic piezospintronics.

**3.1 Relevant ideas**

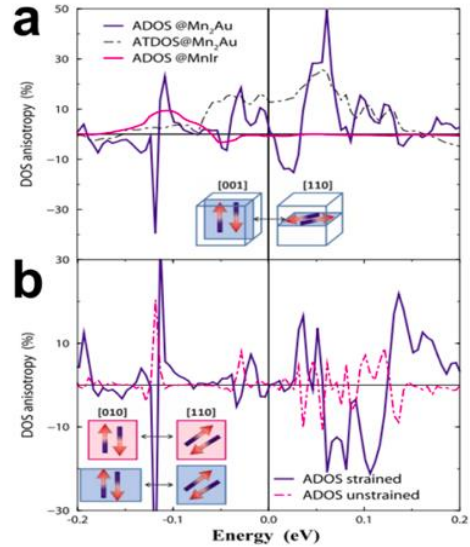

**Figure 1.** Theoretical calculations and simulations. a) Density of states (DOS) anisotropies for the hard [001] and easy [110] axes of $Mn_2Au$, and the hard [001] and easy [100] axes of MnIr. b) In-plane DOS anisotropies for staggered moment aligned along the easy axis [010] of the strained $Mn_2Au$ crystal and along the easy axis [110] of the unstrained $Mn_2Au$. Reproduced with permission.[68] Copyright 2010, American Physical Society.

In 2010, Shick *et al*. revealed strong lattice-parameter-dependent magnetic anisotropies of the ground-state energy, chemical potential, and density of states (**Figure 1**) in metallic antiferromagnets $Mn_2Au$ and MnIr via first-principles density-functional theory calculations and suggested that only a fraction of a percent strain could reorient the spin axis in these antiferromagnets.[68] Later in 2016, Plekhanov *et al*. theoretically examined the possibility of controlling magnetic anisotropy of antiferromagnetic $Mn_2Au$ films via piezoelectric strain generated from $BaTiO_3$ substrates.[69] Afterwards in 2017, Sapozhnik *et al*. experimentally demonstrated the manipulation of the spin axis in an antiferromagnetic $Mn_2Au$ film by a 0.1% tensile strain that was applied via a miniature vacuum compatible stress device by bending a substrate and the $Mn_2Au$ thin film, which supports the strain control in antiferromagnetic spintronics.[70] Sapozhnik *et al*. also suggested that by using piezoelectric substrates, electric-field-driven antiferromagnetic switching could occur in Mn$_2$Au thin films,[70] which would further reduce ohmic losses compared with the current-induced Néel spin-orbit torques in semiconducting CuMuAs.[14] More recently, Chen *et al.* grew Mn$_2$Au thin films on ferroelectric PMN-PT substrates and realized the spin axis manipulation through piezoelectric strain.[71]

On the other hand, motivated by the piezoelectric strain controlled magnetic phase transition in FeRh[60,72] and the FeRh-based antiferromagnetic memory,[11] we started to work on some antiferromagnetic materials whose spin structures are correlated to lattice structures at Oak Ridge National Laboratory. For example, MnPt is a tetragonal collinear antiferromagnet and undergoes a spin-flip transition from its *c* axis to *a* axis while its lattice expands during heating.[73] We were thinking whether we could fabricate high-quality epitaxial MnPt films onto piezoelectric substrates and then manage to manipulate its spin axis by piezoelectric strain as stated in the introduction part of the relevant work.[74] Once this could be realized, the spin axis rotation would be conveniently characterized by the anisotropic magnetoresistance.

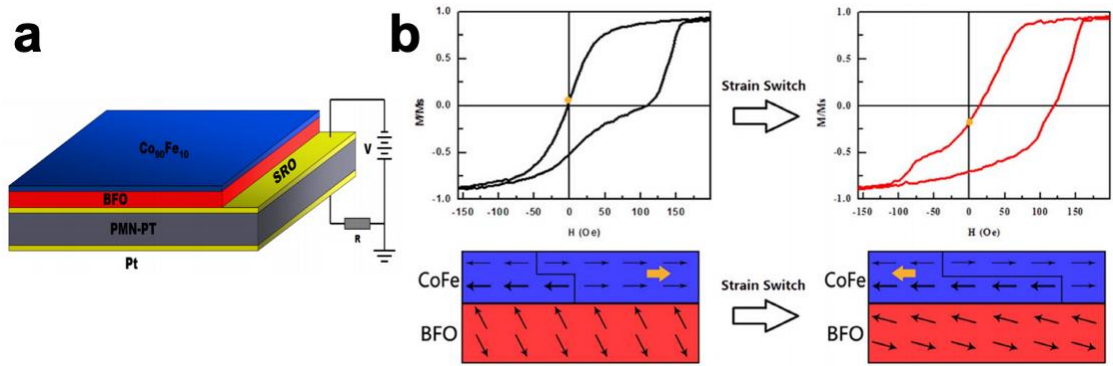

**Figure 2.** a) Schematic diagram of the structure Co$_{90}$Fe$_{10}$ (5 nm)/BiFeO$_3$ (70 nm)/SrRuO$_3$ (5 nm)/PMN-PT/Pt. The polarization voltage was applied between the SrRuO$_3$ (SRO) and Pt electrode in order to avoid directly applying an electric field to the BiFeO$_3$ (BFO) layer. b) Expected BFO antiferromagnetic states under the control of strain. The sketch shows the situation of interface spins at the yellow dots of the *M–H* curves. The yellow arrows indicate the remanent polarization partially changed from positive to negative. Reproduced with permission.[76] Copyright 2015, Springer Nature.

In addition, there are many reports on the piezoelectric control of exchange bias in ferromagnetic/antiferromagnetic bilayer systems. Some of studies attribute the modulation of the exchange bias to the changes of ferromagnetic layers, for example, in the CoO/Co/BaTiO$_3$ heterostructure.[75] There are reports where the exchange bias change was indeed ascribed to the manipulation of the antiferromagnetic spin axis by piezoelectric strain. For example, as early as 2014, Wu *et al*. concluded that the strain generated from PMN-PT substrates changes the spin orientation of antiferromagnetic BiFeO$_3$ and consequently modulates the exchange bias in the Co$_{90}$Fe$_{10}$/BiFeO$_3$/SrRuO$_3$/PMN-PT heterostructure (**Figure 2**).[76] This is indeed rather creative as most of previous studies did not correlate the piezoelectric strain with antiferromagnetic spins at all. Later, Zhang *et al*. identified the piezoelectric strain modulation of antiferromagnetic spins of NiO in Ni/NiO/PMN-PT heterostructures as well and proposed that it could be very useful for antiferromagnetic spintronics.[77] Interestingly, in the NiO/Ni/PMN-PT system, Domann *et al*. identified the modulation of both antiferromagnetic and ferromagnetic spins by the piezoelectric strain in different strain range.[78] As these studies are mainly based on antiferromagnetic oxide insulators such as BiFeO$_3$ and NiO, it is thus difficult to directly exploit them alone for antiferromagnetic spintronics devices. However, by virtue of the inverse spin Hall effect and an additional heavy metal layer such as Pt, the piezoelectric manipulation of spins in antiferromagnetic insulators could be very useful for device applications.[79]

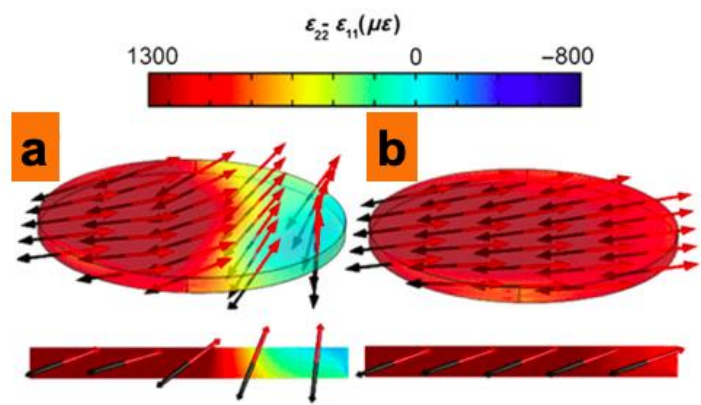

**Figure 3.** The strain (3D color plot) and spin states (black and red arrows) of an antiferromagnetic disk in the composite are plotted for different time. a) At *t* = 71.5 ps, the wave front of the acoustic excitation reaches about halfway across the disk, and the sublattice moments behind the wave front switch in plane, whereas the moments ahead of the wave front do not. b) At *t* = 102.5 ps, the strain propagates across the disk and switches it completely. Reproduced with permission.[80] Copyright 2018, American Physical Society.

Very recently in March of 2018, Barra *et al.* predicted the piezoelectric strain induced antiferromagnetic spin axis switching (**Figure 3**) via a fully coupled finite-element model incorporating micromagnetics, elastodynamics, and piezoelectricity.[80] They theoretically revealed that such a switching can occur at a THz frequency and the switching energy is less than 0.5 fJ/bit, which is much more efficient than other approaches. The numerical simulation in this work is applicable to general antiferromagnetic metals.

**3.2 Realization of piezospintronic effects in antiferromagnetic metals**

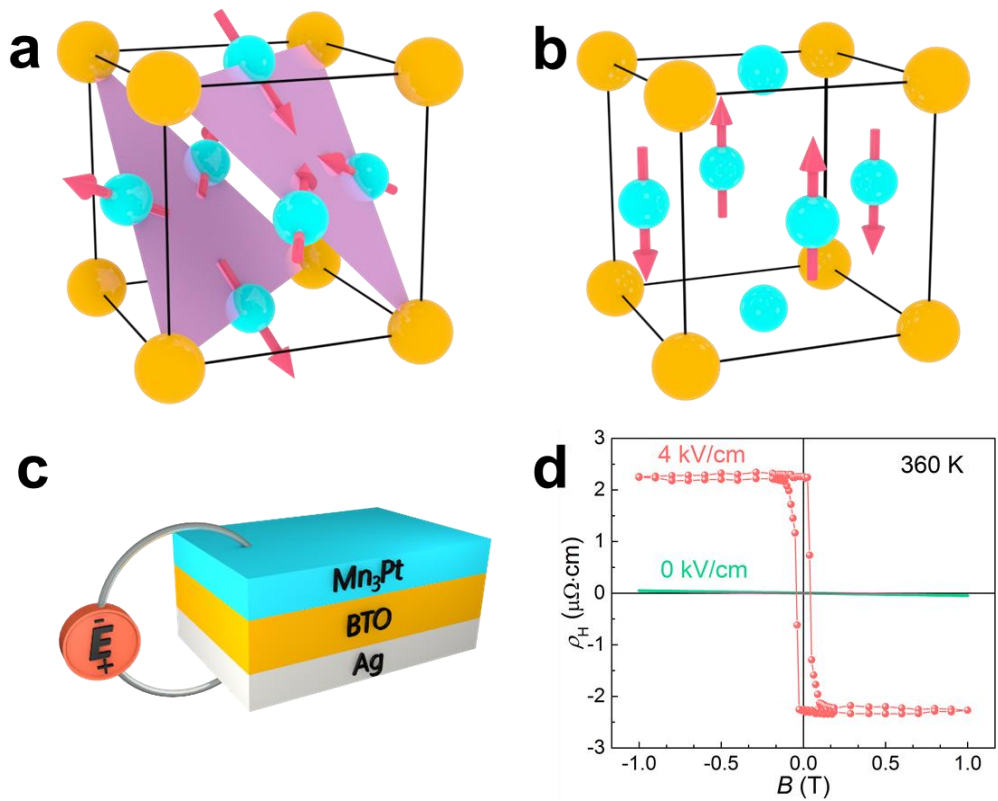

**Figure 4.** Piezoelectric switching of the antiferromagnetic spin structure in $Mn_3Pt$. a,b) Schematics of low-temperature non-collinear and high-temperature collinear antiferromagnetic phase of $Mn_3Pt$, respectively. Reproduced with permission.[58] Copyright 2018, Wiley. c) Schematic of a 20 nm $Mn_3Pt/BaTiO_3$ heterostructure with an electric field $E$ applied perpendicular to the $BaTiO_3$ substrate. d) Hall effect of the $Mn_3Pt$ film under zero electric field and $E$ = 4 kV/cm at 360 K. Reproduced with permission.[81] Copyright 2018, Springer Nature.

From power consumption point of view, piezoelectric strain control of antiferromagnetic spins in metals is highly desired as the Joule heating during information readout could be further suppressed. Almost at the same time of Barra *et*

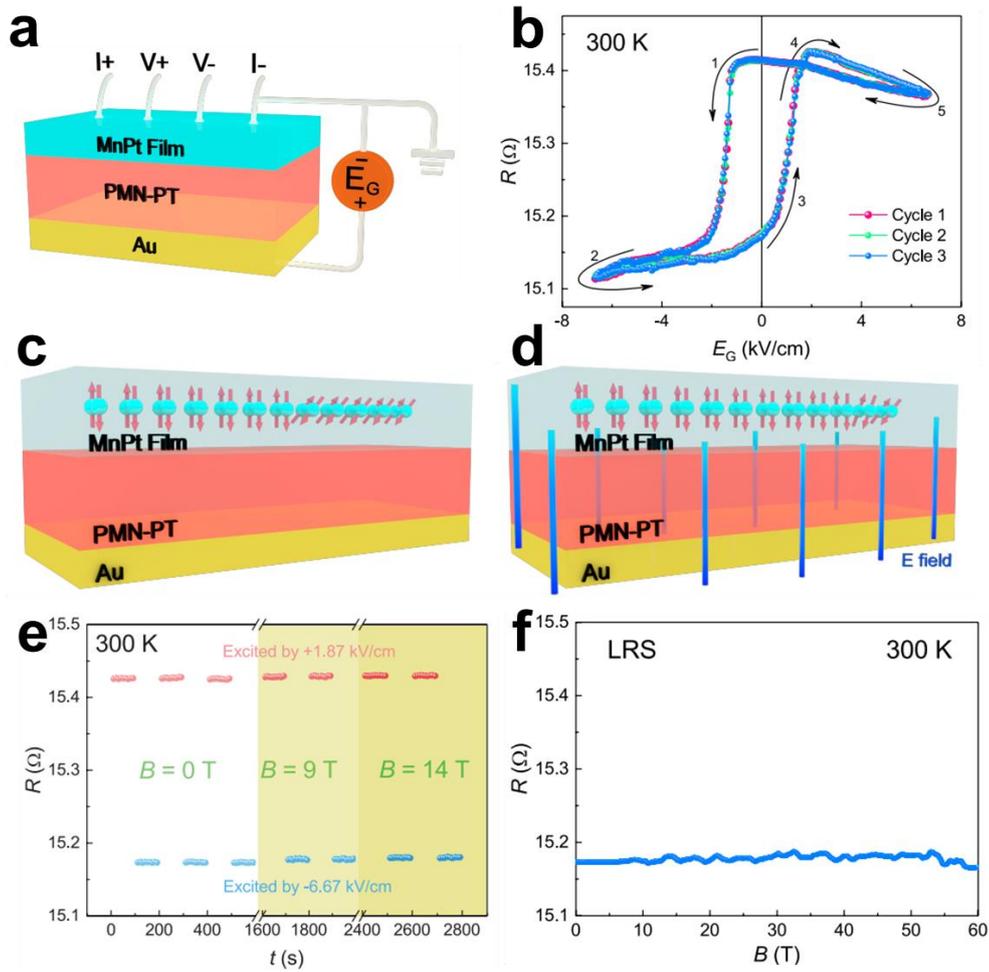

**Figure 5.** A piezoelectric strain-controlled antiferromagnetic memory insensitive to magnetic fields. a) Schematics of electric-field ($E_G$) gating geometry based on a 37-nm-thick MnPt/PMN-PT heterostructure. b) $E_G$-dependent resistance of the MnPt film. c,d) Schematics of the antiferromagnetic spin axis distribution at the high-resistance and low-resistance states, respectively. e) The high-resistance and the low-resistance states realized by pulses of $E_G$ = +1.87 and −6.67 kV/cm at room temperature, respectively, under 0, 9 and 14 T. The area highlighted in yellow on the right shows the existence of a 9 or 14 T magnetic field. f) Magnetic field dependence of the low-resistance state (LRS) under a pulsed magnetic field up to 60 T at room temperature. Reproduced with permission.[87] Copyright 2019, Springer Nature.

*al.*'s numerical prediction, we experimentally realized the reversible piezoelectric strain modulation of the antiferromagnetic spin structures in epitaxial thin films of Mn$_3$Pt, which is very similar to FeRh but undergoes a first-order antiferromagnetic-antiferromagnetic transition from a low-temperature non-collinear phase to a high-temperature collinear phase.[81] Due to the intense interest in Berry phase physics in current condensed matter physics, the anomalous Hall effect in non-collinear antiferromagnets was theoretically predicted by Chen *et al.* in 2014[82] and experimentally confirmed by Nakatsuji *et al*. shortly in 2015.[83] In our work, we used the anomalous Hall effect as an electric probe to the antiferromagnetic spin structure (**Figure 4**). We successfully demonstrated that around the phase transition temperature of $Mn_3Pt$ its antiferromagnetic spin structure is rather sensitive to external piezoelectric strain and can switch the two antiferromagnetic phases back and forth. The main physics is the free energy competition of the two phases under tetragonal distortion of the lattice induced by the piezoelectric strain, which has been emphasized in our earlier publication.[61] As the piezoelectric strain approach is of high energy efficiency, it could be important for the development of antiferromagnetic spintronics.[84-86]

Based on this work, we further focused on the intermetallic MnPt, which is a collinear antiferromagnet and has a high Néel temperature of ~975 K. By integrating textured MnPt films on ferroelectric PMN-PT substrates, we observed two non-volatile resistance states at zero electric field (**Figure 5**a & b). The electroresistance was identified to originate from the rotation of the spin axis of the (101)-oriented grains by the biaxial piezoelectric strain (**Fig. 5**c & d). Owing to the absence of a large current that, for example, is used in the spin-orbit torque approach, the information writing process of this piezoelectric strain controlled memory is very robust under strong magnetic fields (**Fig. 5**e), while the information writing process of the current-operated antiferromagnetic memory is largely affected by a magnetic field of 12 T.[14] Due to strong antiferromagnetic coupling in MnPt, the strain-induced resistance state is stable even at an ultrahigh pulsed magnetic field of 60 T (**Fig. 5**f). Consequently, the combination of the high Néel temperature and the piezoelectric strain approach pushes the insensitivity of antiferromagnets to magnetic fields to its limit in this piezoelectric strain-controlled antiferromagnetic memory.[87] This emphasizes the advantage of piezoelectric strain in efficiently controlling antiferromagnetism.[88] In addition, piezoelectric control of antiferromagnetically coupled spins has been theoretically proposed in honeycomb two-dimensional antiferromagnets with inversion symmetry[89] and experimentally observed in $Mn_3N$[90] and antiperovskite $Mn_3NiN$.[91]

The strain-induced antiferromagnetic spin axis rotation in $L1_0$-type collinear antiferromagnets MnX (X = Pt, Ir, Pd, Rh, Ni) has been recently examined by Park *et al.* through theoretical calculations.[92] They found that a small amount of strain can effectively switch the spin axes of all these materials by 90°. Upon applying external strain, the spin axis rotates within the basal plane for MnIr, MnRh, MnNi and MnPd while it rotates between out of plane and in plane for MnPt, which is consistent with our experimental results.[87]

As emphasized before, the piezoelectric strain from piezoelectric substrates can uniformly work through entire thin films grown on them. Accordingly, we even realized a room-temperature antiferromagnetic tunnel junction via the piezoelectric strain approach and obtained a large tunneling anisotropic magnetoresistance of 11.2% (**Figure 6**),[87] which is close to that of the first room-temperature ferromagnetic tunnel junction achieved by Moodera *et al.* in 1995.[93] Therefore, this progress signifies significant memory application potential of piezoelectric-strain-controlled antiferromagnetic tunnel junctions. We also notice that the similar piezoelectric strain approach has been applied to ferromagnetic tunnel junctions by Chen *et al.* in their recent work.[94]

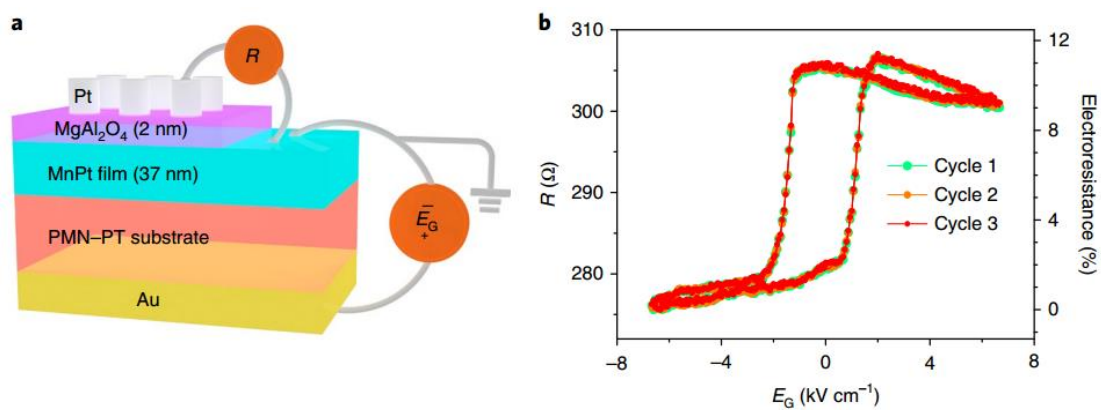

**Figure 6.** A Piezoelectric strain-controlled room-temperature antiferromagnetic tunnel junction.

a) Schematic of the junction structure built on PMN–PT and the measurement geometry. b) Electric-field ($E_G$) dependence of the two-probe tunnelling resistance ($R$) for a Pt(10)/MgAl$_2$O$_4$(2)/MnPt(37)/PMN–PT (thickness in nm) device. Reproduced with permission.[87] Copyright 2019, Springer Nature.

Single-phase multiferroic materials such as BiFeO$_3$ and Cr$_2$O$_3$[58] which involve antiferromagnetic order and ferroelectric order are capable of controlling antiferromagnetic spins via electric fields as well, but they are typically oxide insulators and thus the electric signal readout is not applicable for them alone. In addition, the magnetoelectric coupling coefficients are rather weak compared with artificial multiferroic heterostructures.[58]

Last but not the least, exotic hidden spin and orbital polarizations due to the lack of the local inversion symmetry at atomic sites could play an important role in antiferromagnetic piezospintronics.[95,96] For example, Ryoo and Park found that the hidden orbital polarization in nonmagnetic materials such as diamond, silicon, and germanium could be manifested as real antiferromagnetic magnetization by current when they are under strain that breaks the symmetry of the original crystals. As a prototypical material, it was shown that silicon under a 2% strain could exhibit a magnetoelectric coupling strength comparable to Cr$_2$O$_3$.[96]

To briefly summarize, all the above studies illustrate that piezoelectric strain can behave as a powerful, rather energy-efficient, and multifunctional handle to harnessing spins in antiferromagnets which are difficult to rotate by magnetic fields. Therefore, antiferromagnetic piezospintronics could soon become a rather stimulating area. In this subfield, antiferromagnets' insensitivity to magnetic fields has been already verified by our work.[87] However, the ultrafast switching dynamics as numerically simulated[80] is still pending experimental realization. Moreover, endurance properties of piezoelectric materials are important for realizing reliable pizeospintronic devices. Finally, the spins could be correlated to other electrical, magnetic, optical, thermal, acoustic, mechanical[97] and even chemical properties of antiferromagnetic materials, and thus antiferromagnetic piezospintronics can, in principle, be linked with diverse research fields in an ultralow energy-consuming manner.

**Acknowledgements**

Z.Q.L. acknowledges financial support from the National Natural Science Foundation of China (NSFC Grant No. 51822101, 51861135104, 51771009, & 11704018)